# New developments of the ZZ Ceti instability strip: The discovery of eleven new variables*

B. G. Castanheira[1,2], S. O. Kepler[1], S. J. Kleinman[3], A. Nitta[3], and L. Fraga[4]

[1] *Departamento de Astronomia, Universidade Federal do Rio Grande do Sul, Av. Bento Gonçalves 9500*

*Porto Alegre 91501-970, RS, Brazil*

[2] *Institut für Astronomie, Türkenschanzstr. 17, A-1180 Wien, Austria*

[3] *Gemini Observatory, Northern Operations Center, 670 North A'ohoku Place, Hilo, HI 96720, USA*

[4] *Southern Observatory for Astrophysical Research, Casilla 603, La Serena, Chile*

3 July 2010

**ABSTRACT**

Using the SOAR 4.1 m telescope, we report on the discovery of low amplitude pulsations for three stars previously reported as Not–Observed–to–Vary (NOV) by Mukadam et al. (2004a) and Mullally et al. (2005), which are inside the ZZ Ceti instability strip. With the two pulsators discovered by Castanheira et al. (2007), we have now found variability in a total of five stars previously reported as NOVs. We also report the variability of eight new pulsating stars, not previously observed, bringing the total number of known ZZ Ceti stars to 148. In addition, we lowered the detection limit for ten NOVs located near the edges of the ZZ Ceti instability strip. Our results are consistent with a pure mass dependent ZZ Ceti instability strip.

## 1 INTRODUCTION

The ZZ Ceti stars (or DAVs) are pulsating white dwarfs with hydrogen-dominated atmosphere (DAs) and are observed in a narrow instability strip, between 10 800 and 12 300 K [e.g.

---

* Based on observations at the SOuthern Astrophysical Research telescope, a collaboration between CNPq-Brazil, NOAO, UNC and MSU

(Bergeron et al. 2004; Mukadam et al. 2004a)], with a small dependency on mass (Giovannini et al. 1998, e.g.).

Mukadam et al. (2004a) and Mullally et al. (2005) reported roughly twenty stars as Not–Observed–to–Vary (NOV) inside the instability strip, with amplitude limits of the order of 4 mma, close to the amplitude of previously known smallest amplitude pulsators. Considering only the ZZ Ceti stars discovered by the Sloan Digital Sky Survey (SDSS), Mukadam et al. (2004b) computed the likelihood that the instability strip was pure to be $\sim 0.044\%$. Their result disagreed both with previous observations and with pulsation models which predict pulsation as a normal phase in the cooling of all white dwarfs. Perhaps additional physical processes should be included in the models (e.g. strong magnetic fields), they theorized, to match the observations.

Gianninas, Bergeron, & Fontaine (2005, 2006) proposed that the instability strip is pure, based on their 100% success rate in predicting variability for DA stars, if temperature and surface gravity determinations are reliable, i.e., if these quantities are derived from S/N>60 spectra. Their argument was based on the study of the brightest sample of candidate pulsators, from the catalog of McCook & Sion (1999). Gianninas, Bergeron, & Fontaine (2005, 2006) claimed that the uncertainties in temperature and mass for the dimmer stars in the SDSS sample used by Mukadam et al. (2004a) were large enough to scatter pulsators outside the instability strip and constant stars, inside. Kepler et al. (2006) re-observed four stars with $T_{\rm eff} \sim 12\,000$ K from the SDSS sample with GMOS at the Gemini 8 m telescope; they took spectra with S/N⩾60. Fitting these high S/N spectra, they estimated the real uncertainties in the low S/N SDSS spectra fits of Kleinman et al. (2004) and Eisenstein et al. (2006) are larger than the quoted internal uncertainties by 60% in temperature and a factor of 4 in $\log g$. However, the main component of their fit disagreement was systematic, with an average difference from the SDSS catalog measurements (Kleinman et al. 2004; Eisenstein et al. 2006) in temperature of 320 K, systematically lower, and in $\log g$ of 0.24 dex, systematically larger. Because the differences are systematic, it would appear that low S/N of the SDSS spectra are not the main explanation for the possibly contaminated instability strip. On the other hand, if the uncertainties in $T_{\rm eff}$ are of the order of 300 K, there should be some scatter of pulsators out of the strip and constant stars inside, as the strip is only $\sim 1500$ K wide.

The present paper is a continuation of the effort to determine whether the ZZ Ceti instability strip is a normal evolutionary stage in the white dwarf evolution or not by exploiting the real purity of the instability strip. We seek to understand if the asteroseismological measurements of pulsating white dwarfs can be applied to white dwarfs in general, which represent the endpoint of the evolution of more than 95% of all stars. In this paper, we report the variability of three ZZ Ceti stars previously classified as NOVs inside the instability strip and seven previously unobserved variables. In our searches, we also lowered the detection limit for ten NOVs near the edges of the ZZ Ceti instability strip. Since we have discovered low amplitude pulsations in *every* NOV inside the instability strip we have observed, our observational evidence for a pure ZZ Ceti instability strip. However, we will only be able to claim that the ZZ Ceti instability strip is truly pure or not when we lower the detection limits for variability of all stars within the boundaries of the strip. It is of the same importance to have more accurate $T_{\text{eff}}$ and $\log g$ determinations for all stars near the edges and inside the instability strip.

## 2 OBSERVATIONS AND DATA REDUCTION

We are looking for pulsators among the white dwarfs discovered with the SDSS (Abazajian et al. 2009). Kleinman et al. (2004) describes the fitting process for all SDSS white dwarfs. Briefly, they combined the SDSS photometry, along with re-fluxed whole-spectrum model fits, to obtain $T_{\text{eff}}$ and $\log g$ measurements, with a fitting program called *autofit*. For our work, we used the same version of autofit as in Eisenstein et al. (2006), but with a new, extended model grid and with SDSS DR7 reductions. The details of these fits will be described in a coming paper on DR7 SDSS white dwarfs. We included in our candidate list stars with temperatures comparable to the current observed ZZ Ceti instability strip. We also observed previously-observed NOVs that appear in the instability strip with detection limits above 1 mma (Castanheira et al. 2007).

We observed our targets with the 4.1 m SOAR telescope, in Chile, using the SOAR Optical Imager, a mosaic of two EEV 2048×4096 CCDs, thinned and back illuminated, with an efficiency around 73% at 4 000 Å, at the Naysmith focus. The observations were carried out in service mode by the SOAR staff of Brazilian Resident Astronomers. The integration times were 30 s. We used fast readout mode with the CCDs binned 4×4 to decrease the readout+write time to 6.4 s and still achieve 0.354"/pixel resolution. All observations were obtained with a Bessel *B* filter to maximize the amplitude and minimize the red fringing. Tables 1 and 2 present the journal of observations.

| Star | Run start (UT) | $t_{\rm exp}$ (s) | $\Delta T$ (h) | # Points |
|---|---|---|---|---|
| SDSS J003719.12+003139.3 | 2007-11-12 00:33 | 30 | 2.0 | 200 |
| | 2008-09-23 05:19 | 30 | 2.0 | 199 |
| | 2008-10-02 05:09 | 30 | 2.1 | 201 |
| | 2008-10-25 00:56 | 30 | 2.0 | 200 |
| | 2008-10-26 00:02 | 30 | 2.1 | 199 |
| | 2008-10-27 00:15 | 30 | 2.0 | 200 |
| SDSS J004345.78+005549.9 | 2006-11-08 01:46 | 30 | 2.7 | 270 |
| SDSS J005047.60-002316.9 | 2005-12-03 00:59 | 30 | 4.0 | 399 |
| | 2005-12-04 00:34 | 30 | 4.2 | 300 |
| | 2005-12-05 00:44 | 30 | 4.3 | 422 |
| | 2007-10-02 03:33 | 30 | 2.9 | 286 |
| | 2007-10-03 02:47 | 30 | 2.0 | 200 |
| SDSS J012234.68+003025.8 | 2007-10-02 06:43 | 30 | 2.5 | 248 |
| | 2007-10-04 05:19 | 30 | 1.5 | 149 |
| | 2007-11-09 01:20 | 30 | 3.0 | 300 |
| SDSS J012950.44-101842.0 | 2009-08-23 07:12 | 30 | 2.8 | 295 |
| | 2009-08-25 07:56 | 30 | 2.3 | 250 |
| SDSS J015259.18+010017.7 | 2008-10-25 03:03 | 30 | 2.0 | 196 |
| | 2008-10-26 02:45 | 30 | 2.0 | 200 |
| | 2008-10-27 02:25 | 30 | 2.0 | 200 |
| SDSS J025709.00+004628.1 | 2008-10-27 04:36 | 30 | 2.0 | 201 |
| | 2008-11-05 02:36 | 30 | 2.0 | 201 |
| SDSS J030153.81+054020.0 | 2009-08-14 08:05 | 30 | 2.0 | 201 |
| SDSS J032302.86+000559.6 | 2008-09-23 07:29 | 30 | 2.0 | 200 |
| | 2008-10-27 04:36 | 30 | 2.2 | 196 |
| SDSS J033648.34-000634.4 | 2008-10-02 07:46 | 30 | 1.8 | 174 |
| | 2008-10-04 06:33 | 30 | 2.5 | 241 |
| SDSS J034504.21-003613.4 | 2005-12-09 01:15 | 30 | 4.5 | 444 |
| | 2006-01-05 00:49 | 30 | 4.4 | 393 |
| | 2008-10-26 07:07 | 30 | 1.6 | 155 |
| SDSS J082239.43+082436.7 | 2008-02-08 03:46 | 30 | 2.0 | 200 |
| | 2008-02-09 00:59 | 30 | 2.1 | 219 |
| | 2008-03-10 01:02 | 30 | 2.0 | 200 |
| | 2008-03-12 02:37 | 30 | 2.0 | 200 |
| SDSS J092511.60+050932.4 | 2008-01-27 05:43 | 30 | 2.0 | 199 |
| | 2008-01-30 05:53 | 30 | 2.0 | 201 |
| SDSS J095936.96+023828.4 | 2008-02-20 03:06 | 30 | 2.2 | 200 |
| | 2008-03-10 03:23 | 30 | 2.1 | 200 |
| SDSS J110525.70-161328.5 | 2010-01-30 04:13 | 30 | 1.8 | 200 |
| | 2010-02-11 05:45 | 30 | 2.6 | 285 |
| SDSS J113604.01-013658.1 | 2007-03-19 01:02 | 30 | 2.5 | 210 |
| | 2007-03-24 06:23 | 30 | 2.1 | 210 |
| SDSS J133831.74-002328.0 | 2007-03-26 03:38 | 30 | 2.1 | 211 |
| | 2007-04-20 02:36 | 30 | 2.1 | 210 |
| | 2007-05-14 00:40 | 30 | 2.1 | 209 |

**Table 1.** Journal of observations for the ZZ Ceti candidates using the 4.1 m SOAR telescope. $\Delta T$ is the total length of each observing run and $t_{\rm exp}$ is the integration time of each exposure.

| Star | Run start (UT) | $t_{\exp}$ (s) | $\Delta T$ (h) | # Points |
|---|---|---|---|---|
| SDSS J143249.10+014615.5 | 2007-04-20 07:36 | 30 | 1.8 | 175 |
| | 2007-04-21 04:33 | 30 | 2.2 | 215 |
| | 2007-05-12 04:25 | 30 | 2.1 | 210 |
| SDSS J214723.73-001358.4 | 2008-08-04 03:47 | 30 | 2.0 | 199 |
| | 2008-08-05 04:26 | 30 | 2.0 | 201 |
| SDSS J220915.84-091942.5 | 2007-05-13 07:52 | 30 | 2.1 | 210 |
| | 2007-06-17 06:02 | 30 | 4.0 | 400 |
| SDSS J232659.22-002347.8 | 2007-09-20 05:51 | 30 | 1.1 | 104 |
| | 2008-08-04 05:55 | 30 | 2.0 | 200 |
| | 2008-09-23 02:60 | 30 | 2.0 | 198 |
| | 2008-10-02 00:04 | 30 | 2.1 | 200 |
| SDSS J234141.61-010917.2 | 2008-10-03 01:16 | 30 | 2.2 | 202 |
| | 2008-10-04 02:50 | 30 | 2.0 | 200 |

**Table 2.** Continuation of Table 1.

We reduced the data using *hsp* (high speed photometry) scripts, developed by Antonio Kanaan for IRAF, with weighted apertures, for time-series photometry (Kanaan et al. 2005). We extracted light curves of all bright stars that were observed simultaneously in the field. Then, we divided the light curve of the target star by either the brightest comparison or a sum of the light curves of the comparison stars to minimize effects of sky and transparency fluctuations. We chose our aperture size by optimizing the noise in the resulting Fourier transform.

As an objective criterion for determining which peaks are real in the Fourier transform, we determined a power amplitude limit such that a peak exceeding this limit has a 1/1000 probability of being due to noise (false alarm probability or FAP), following the general description of Scargle (1982) and application for white dwarfs of Kepler (1993). For each light curve, we calculated the ratio $P_0/\langle P \rangle = \ln(\frac{1}{1000*N})$, where $P_0$ is the power amplitude of a peak, $\langle P \rangle$ is the average in the power spectrum, and $N$ is the number of independent samples.

We observed most targets at two separate times, each for of order two hours, to look for coherent signals in the light curves, as listed in Tables 1 and 2. If a pulsation was detected, we also checked if smaller peaks in the Fourier transform were intrinsic variations of the star or aliases due to the spectral window. For this, we subtracted from the original light curve the sinusoid representing the highest amplitude peak, i.e., pre-whitening. The subtracted sine curve had the same amplitude, period, and phase information as the peak selected in the Fourier transform. After this subtraction, we re-calculated the Fourier transform and the new noise level. We continued pre-whitening until the highest remaining peak has FAP greater than 1/1000, as described in the previous paragraph.

| Star | $T_{\rm eff}$ (K) | $\log g$ | Mass ($M_\odot$) | g (mag) | Period (s) | Amplitude (mma) |
|---|---|---|---|---|---|---|
| SDSS J004345.78+005549.9 | 11820±190 | 7.94±0.10 | 0.58±0.05 | 18.74 | 258.24 | 6.69 |
| SDSS J012234.68+003025.8 | 11800±50 | 7.87±0.02 | 0.54±0.01 | 17.29 | 121.07<br>200.75<br>358.61 | 1.53<br>1.25<br>1.23 |
| SDSS J012950.44-101842.0 | 11910±130 | 8.00±0.03 | 0.61±0.02 | 18.32 | 193.76<br>147.42 | 2.88<br>2.33 |
| SDSS J030153.81+054020.0 | 11470±50 | 8.09±0.03 | 0.66±0.02 | 18.05 | 300.83 | 24.87 |
| SDSS J092511.60+050932.4 | 10880±30 | 8.41±0.02 | 0.87±0.01 | 15.20 | 1127.14<br>1264.29 | 3.17<br>3.05 |
| SDSS J095936.96+023828.4 | 11840±110 | 8.05±0.06 | 0.64±0.04 | 18.15 | 283.41<br>194.68 | 12.95<br>7.23 |
| SDSS J110525.70-161328.5 | 11670±90 | 8.23±0.03 | 0.75±0.02 | 17.54 | 192.66<br>298.25 | 12.09<br>7.09 |
| SDSS J113604.01-013658.1 | 11710±70 | 7.96±0.04 | 0.59±0.02 | 17.84 | 260.79 | 2.45 |
| SDSS J133831.74-002328.0 | 11870±80 | 8.13±0.04 | 0.69±0.02 | 17.09 | 196.93<br>119.72 | 3.97<br>1.75 |
| SDSS J214723.73-001358.4 | 11990±290 | 7.92±0.11 | 0.57±0.06 | 18.98 | 199.77 | 3.88 |
| SDSS J220915.84-091942.5 | 11430±110 | 8.33±0.06 | 0.82±0.04 | 18.93 | 894.71<br>447.94<br>789.31 | 43.94<br>10.80<br>10.37 |

**Table 3.** Observational properties of the new ZZ Ceti stars. $T_{\rm eff}$ and $\log g$ were determined from SDSS spectra.

## 3  NEW ZZ CETI STARS

In Table 3, we list the properties of the new ZZ Ceti stars, discovered in this work. In Figures 1 and 2, we show the Fourier transforms of the previously unobserved ZZ Ceti stars.

Most of the new pulsators are low amplitude, with the exception of SDSS J220915.84-091942.5. This star is a typical red edge pulsator, the cooler end of the instability strip, with high amplitude and long periodicities. For this star, we detected two independent periodicities as well as the first harmonic of the main mode at ∼895 s. Based on observations of other ZZ Ceti stars with similar periodicities and amplitudes, it is reasonable to expect many more modes to be excited in this star, likely to be revealed with longer observing runs.

Among the new pulsators, SDSS J092511.60+050932.4 also pulsates with long periods, but with small amplitude. This star could be an example of a ZZ Ceti on the verge of leaving the instability strip, ceasing to pulsate. There are only a few other stars showing this behavior; Mukadam et al. (2006) and Castanheira & Kepler (2009) discuss this evolutionary phase. The other new pulsators are closer to the blue edge of the instability strip, with low amplitude and short periods.

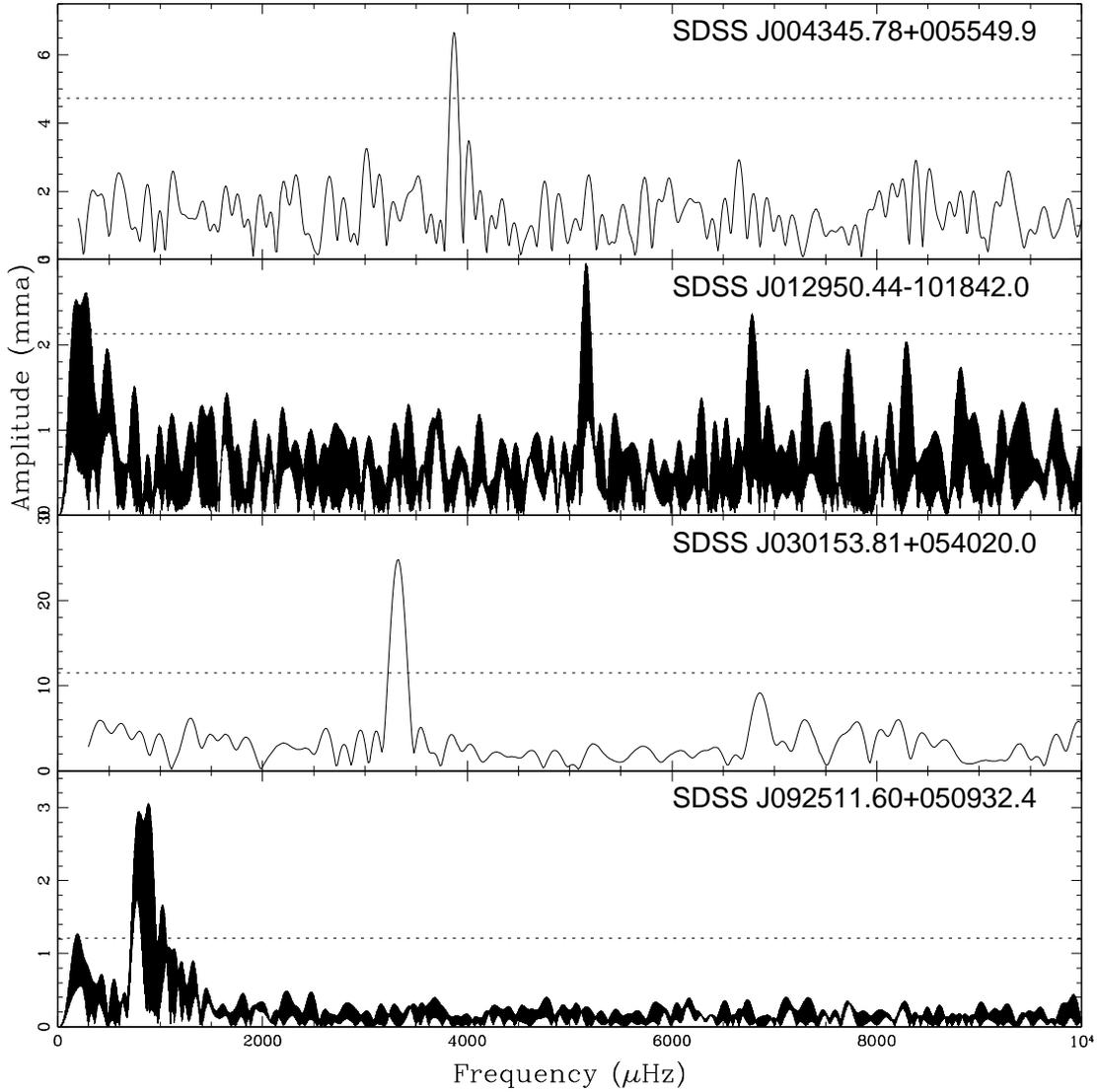

**Figure 1.** Fourier transform of the combined data sets for new ZZ Ceti stars (full line) and the detection limits (dotted line). Note the y-axis is in mma (mili-modulation amplitude) and with a different scale for each star, as the amplitudes are different.

The stars SDSS J012234.68+003025.8, SDSS J113604.01-013658.1, and SDSS J133831.74-002328.0 were previously reported as NOV2, NOV2, and NOV4 by Mukadam et al. (2004a) and Mullally et al. (2005), but are in fact low amplitude pulsators. In Figure 3 we show the Fourier transform of the combined light curves for these stars. Our observations achieved a lower noise level, revealing that these stars are low amplitude pulsators. All of them pulsate

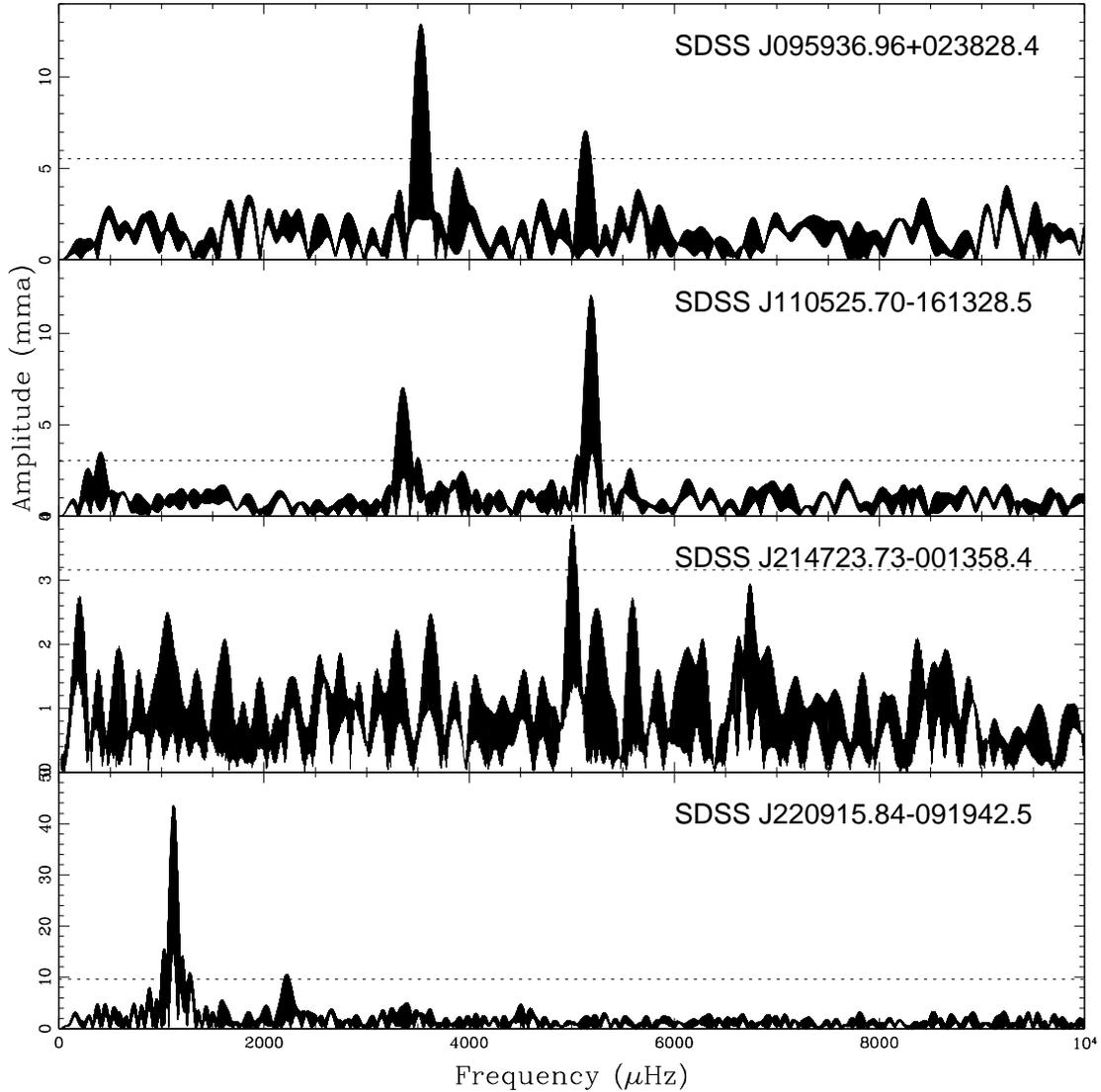

**Figure 2.** Fourier transform of the combined data sets for new ZZ Ceti stars (full line) and the detection limits (dotted line). Note the y-axis is in mma (mili-modulation amplitude) and with a different scale for each star, as the amplitudes are different.

with short periods, typical of blue edge stars, consistent with our temperature determinations.

As expected, all NOVs that we are re-classifying as variables pulsate with low amplitudes below their previous detection limits. All of them have pulsations of short periods, typical of blue edge stars. Even with all the improvements in the models and in the fits since Kleinman et al. (2004) and Eisenstein et al. (2006) determination, the latest temperature and gravity for these pulsators are consistent with the previous values, i.e., inside the instability strip.

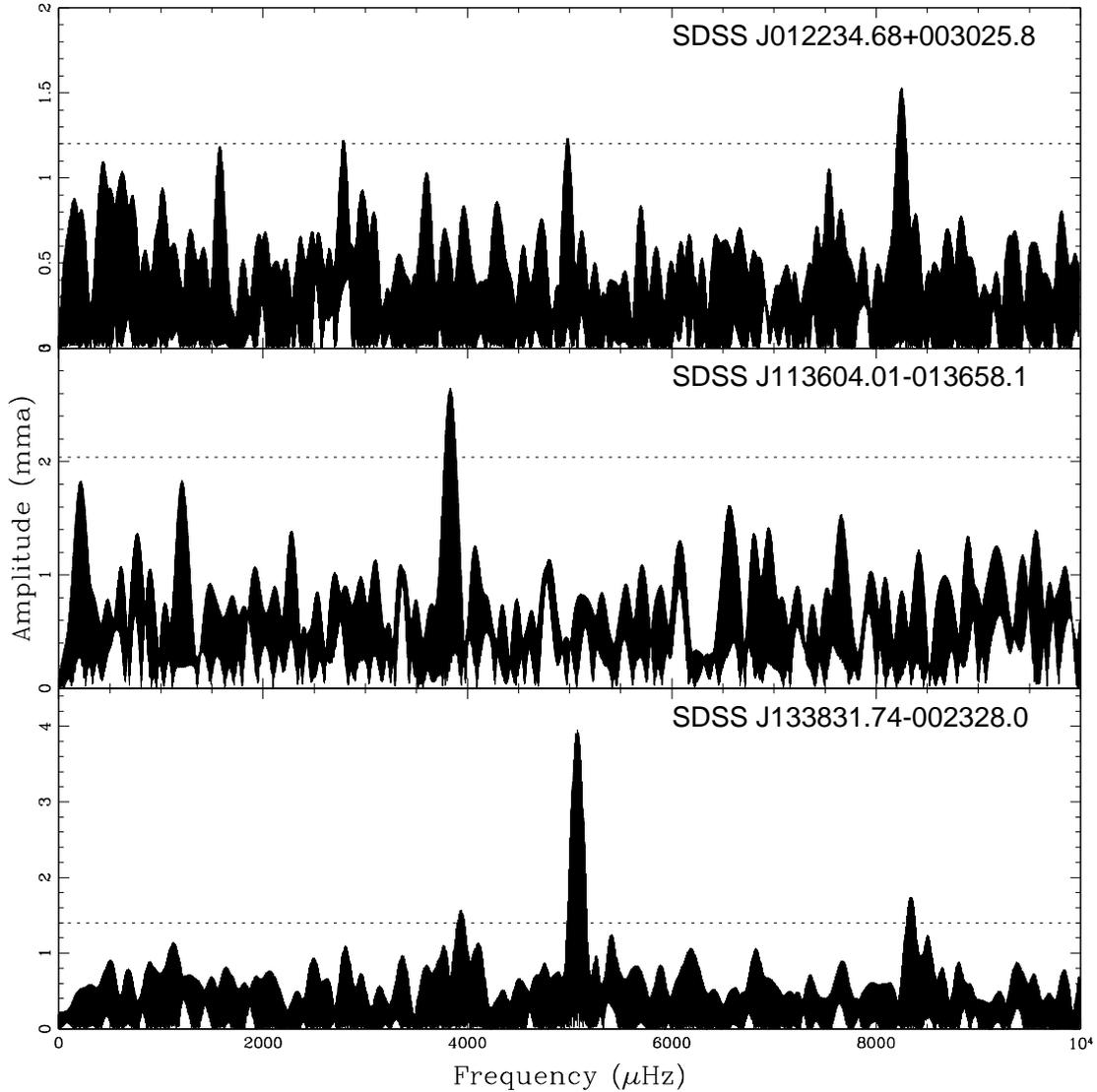

**Figure 3.** Fourier transform of the new ZZ Ceti stars, previously classified as NOVs (full lines) and the detection limits (dotted lines).

## 4 SEISMOLOGY OF THE NEW ZZ CETI STARS

Even though we detected only a few modes for each star, mainly due to our short observing runs, we used the periodicities and amplitudes detected for the new ZZ Ceti stars to do a first seismological study, following Castanheira & Kepler (2009), i.e., not probing the core composition but keeping it fixed. As the number of detected modes is small, we mainly wanted to test if the observed periods are consistent with the $T_{\rm eff}$ and $\log g$ from the spectra. Our seismological study is a starting point for future investigations, varying more internal parameters, possible with the detection of more modes. Castanheira & Kepler (2008) show that the core composition and profile can be compensated in the fits by a change in the He layer, as they probe the internal transition layers. We compared the observed periodicities, weighted by amplitude, with seismological models, similar to a $\chi^2$ fit, according to the expression:

$$S = \sum_{i=1}^{n} \sqrt{\frac{[P_{\text{obs}}(i) - P_{\text{model}}]^2 \times w_P(i)}{\sum_{i=1}^{n} w_P(i)}} \qquad (1)$$

where $n$ is the number of observed modes, $w_P = \frac{1}{A^2}$ is the weight given to each mode, and $A$ is the observed amplitude.

We used the spectroscopic determinations of temperature and $\log g$ to guide our searches for the possible families of seismological solutions. For each family of solution, we list in Table 4 the absolute minimum in $S$, which is the best fit, and the values of $\ell$ (the total number of node lines on the stellar surface) and $k$ (the number of nodes in the pulsation eigenfunction along the radial direction).

We started our studies with the star SDSS J012234.68+003025.8. Its main mode is around 120 s, and there are also other two independent modes at 201 s and 359 s. Based on the arguments discussed in section 2.2 of Castanheira & Kepler (2009), the second largest mode, around 200 s, could fit $\ell = 2$. The amplitudes of the other modes are comparable to the amplitude of this mode, therefore, we tested both $\ell = 1$ and 2 for all observed modes. The minimum of the best seismological solution listed in Table 4 has $M_H = 10^{-4} M_\star$.

The stars SDSS J012950.44-101842.0, SDSS J110525.70-161328.3, and SDSS J133831.74-002328.0 have their main modes around 195 s (see Table 3) as well as an additional mode. Once again, we tested both $\ell = 1$ and 2 for the modes, listing the best seismological solutions in Table 4.

Another star with its main mode around 200 s is SDSS J214723.73-001358.4. Different than for the previous analyzed stars, the spectroscopic temperature places this star at the blue edge of the ZZ Ceti instability strip. The minima of the seismological solutions are listed in Table 4, confirming the result of Castanheira & Kepler (2009), reveals a solution with the main mode as $\ell = 1$, consistent with the spectroscopic temperature.

Moving towards the center of the instability strip, the stars SDSS J004345.78+005549.9 and SDSS J113604.01-013658.1 each pulsate with a mode around 260 s. In Table 4, we list the minima of the possible families of seismological solutions compatible with spectroscopic temperature and mass.

| Star | $T_{\rm eff}$ (K) | M ($M_\odot$) | $-\log M_{\rm H}$ | $-\log M_{\rm He}$ | $S$ (s) | Modes ($\ell, k$) |
|---|---|---|---|---|---|---|
| SDSS J012234.68+003025.8 | 11 800 | 0.54 | 4 | 2 | 2.24 | 120.32(2,1), 205.19(1,1), 358.96(2,9) |
| SDSS J012950.44-101842.0 | 11 300 | 0.61 | 6.5 | 2.5 | 0.26 | 146.97(2,2), 193.67(2,3) |
|  | 11 400 | 0.58 | 7.5 | 3.5 | 0.40 | 147.95(2,1), 193.44(2,2) |
| SDSS J110525.70-161328.3 | 11 500 | 0.73 | 8.5 | 2 | 0.24 | 192.40(2,3), 298.25(2,7) |
| SDSS J133831.74-002328.0 | 11 400 | 0.695 | 4 | 2 | 0.15 | 119.07(2,1), 197.02(2,4) |
|  | 11 800 | 0.71 | 8.5 | 2 | 0.16 | 119.36(2,1), 196.78(2,3) |
| SDSS J214723.73-001358.4 | 11 800 | 0.585 | 5.5 | 2 | 0.03 | 199.74(1,1) |
|  | 12 100 | 0.58 | 5.5 | 2.5 | 0.08 | 199.85(1,1) |
| SDSS J004345.78+005549.9 | 12 100 | 0.585 | 6.5 | 2 | 0.004 | 258.24(1,2) |
|  | 12 100 | 0.54 | 6 | 3 | 0.01 | 258.23(1,2) |
|  | 11 700 | 0.62 | 6.5 | 3.5 | 0.06 | 258.30(1,2) |
| SDSS J113604.01-013658.1 | 11 950 | 0.58 | 6.5 | 2 | 0.06 | 260.85(1,2) |
|  | 11 800 | 0.58 | 6 | 3.5 | 0.02 | 260.77(1,2) |
| SDSS J095936.96+023828.4 | 11 500 | 0.655 | 8 | 2 | 0.59 | 193.90(2,3), 282.85(2,6) |
|  | 11 400 | 0.64 | 9.5 | 3 | 0.35 | 196.43(2,2), 286.97(2,5) |

**Table 4.** Absolute minima in $S$ for the possible families of solutions in the seismological analysis of the new ZZ Ceti stars.

With two detected modes, SDSS J095936.96+023828.4 has a main mode around 280 s and a second mode around 195 s (see Table 3). For the same arguments used for other stars, this second mode is probably $\ell = 2$. If we look for solutions where the main mode is $\ell = 1$, the seismological temperature is 500 K hotter than the spectroscopic determination. For this reason and because the amplitudes of the modes are comparable, we also considered the main mode to be $\ell = 2$. The minima of the two families of solutions are in Table 4.

For all the stars listed above, our initial seismological determinations of temperature and mass are compatible with our spectroscopic fit values.

All the other stars that we discovered are closer to the red edge of the instability strip, where the mode spacings are close to asymptotic, making it more difficult to do even a preliminary asteroseismological analysis (Castanheira & Kepler 2009; Bischoff-Kim 2009).

In Figure 4, we plot the comparison of temperature (upper panel) and mass (lower panel) determinations from spectroscopy (y-axis) and seismology (x-axis). Each symbol represents a different star: the open triangle for SDSS J012234.68+003025.8, the open squares for SDSS J012950.44-101842.0, the open circles for SDSS J133831.74-002328.0, the star for SDSS J110525.70-161328.3, the filled triangles for SDSS J214723.73-001358.4, the filled squares for SDSS J004345.78+005549.9, the filled circles for SDSS J113604.01-013658.1, and the asterisks for SDSS J095936.96+023828.4. The dotted line is the 1:1 agreement. The temperatures from seismology are on average lower than from SDSS spectra. The mass determinations from seismology and spectroscopy are in good agreement.

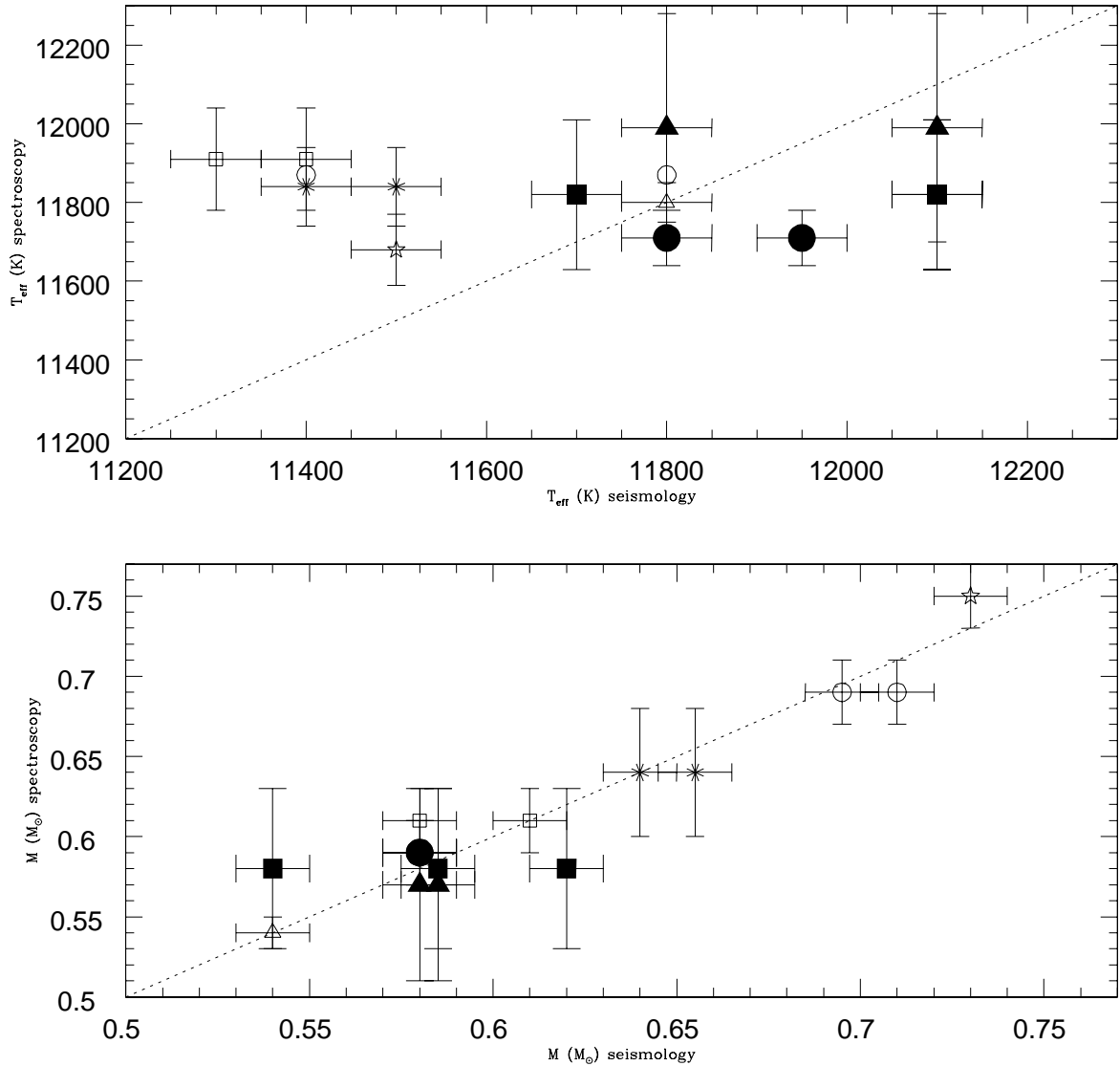

**Figure 4.** Comparison between temperature (upper panel) and mass (lower panel) determinations from spectroscopy (y-axis) and seismology (x-axis). Different symbols were used for different stars: the open triangle for SDSS J012234.68+003025.8, the open squares for SDSS J012950.44-101842.0, the open circles for SDSS J133831.74-002328.0, the star for SDSS J110525.70-161328.3, the filled triangles for SDSS J214723.73-001358.4, the filled squares for SDSS J004345.78+005549.9, the filled circles for SDSS J113604.01-013658.1, and the asterisks for SDSS J095936.96+023828.4. The dotted line represents the 1:1 agreement.

## 5 NEW ZZ CETI INSTABILITY STRIP AND CONSTANT STARS

In the Figure 5, we plot the ZZ Ceti instability strip. The open triangles represent previously known ZZ Ceti stars, while the filled triangles represent our newly discovered pulsators. The circles are the NOVs, with the filled ones being the ones which we have lowered the detection limits and the open ones the stars for which no variability was detected (Mukadam et al.

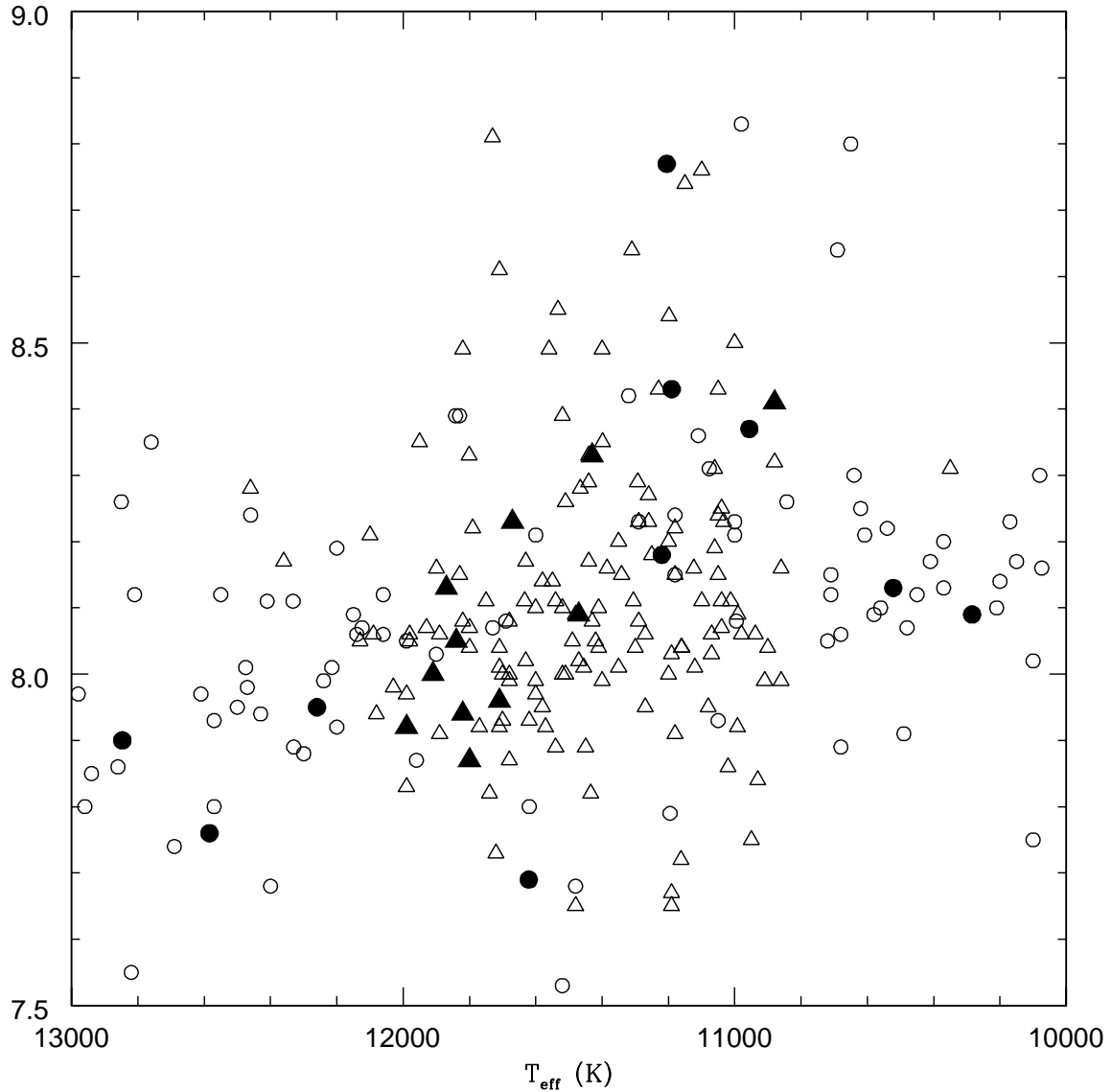

**Figure 5.** New ZZ Ceti instability strip. The full triangles are the ZZ Ceti stars we discovered, the open triangles are the previously known ZZ Ceti stars, the full circles are the NOVs for which we lowered the detection limits, and the open circles are the NOVs from Mukadam et al. (2004a); Mullally et al. (2005); Gianninas, Bergeron, & Fontaine (2005). $T_{\rm eff}$ and $\log g$ are from the spectroscopic determinations, not from seismology.

2004a; Mullally et al. 2005; Gianninas, Bergeron, & Fontaine 2005), but with high amplitude limits. In Table 5 we show the properties and the detection limits of our NOVs.

Although we have re-classified three more stars that are inside the ZZ Ceti instability strip, revealing that they are in fact low amplitude pulsators, there are still more than a dozen NOVs inside the instability strip. The question whether all the remaining NOVs pulsate with amplitudes below the published detection limits remains, since all new low amplitude pulsators have amplitudes smaller than the previous 4 mma average limit. Our observations point towards a pure instability strip, but there is no guarantee that other physical mechanisms cannot shut down pulsations. Therefore, we encourage the search of variability for the stars inside and at the edges of the instability strip previously reported as NOVs, reaching a detection limit of ∼1 mma, before declaring them non-pulsators. For instance, SDSS J034504.21-003613.4 is in the middle of the ZZ Ceti instability strip, but our current detection limit of 3 mma does not exclude this star from our candidate list. However, we caution that once we get our pulsation limits of the stars within the instability strip down to of order 1 mma, we also have to start looking at the stars *outside* the instability strip to the same level.

| Star | $T_{\rm eff}$ (K) | $\log g$ | Mass ($M_\odot$) | g (mag) | New limit |
|---|---|---|---|---|---|
| SDSS J003719.12+003139.3 | 10 960±70 | 8.37±0.05 | 0.84±0.03 | 17.48 | NOV1 |
| SDSS J015259.18+010017.7 | 12 580±60 | 7.76±0.02 | 0.49±0.01 | 16.42 | NOV1 |
| SDSS J025709.00+004628.1 | 12 260±640 | 7.95±0.07 | 0.58±0.04 | 17.39 | NOV1 |
| SDSS J032302.86+000559.6 | 13 450±280 | 7.81±0.03 | 0.52±0.02 | 17.44 | NOV2 |
| SDSS J033648.34-000634.4 | 10 280±60 | 8.09±0.07 | 0.66±0.04 | 17.94 | NOV2 |
| SDSS J034504.21-003613.4 | 11 620±260 | 7.69±0.11 | 0.46±0.05 | 19.00 | NOV3 |
| SDSS J082239.43+082436.7 | 11 190±80 | 8.43±0.05 | 0.88±0.03 | 18.12 | NOV2 |
| SDSS J143249.10+014615.5 | 11 220±70 | 8.18±0.05 | 0.72±0.03 | 17.50 | NOV3 |
| SDSS J232659.22-002347.8 | 10 520±50 | 8.13±0.05 | 0.69±0.03 | 17.49 | NOV2 |
| SDSS J234141.61-010917.2 | 12 850±300 | 7.90±0.08 | 0.56±0.04 | 18.02 | NOV2 |

**Table 5.** Observational properties and detection limits for the NOVs. $T_{\rm eff}$ and $\log g$ were determined from SDSS spectra.

## 6　THE HIGH MASS CANDIDATES

Among the candidate ZZ Ceti stars discovered by SDSS, there are some which have high mass. To truly understand the pulsations in these stars, we need to take into consideration the effects of crystallization in the observed modes, since pulsations cannot propagate into the crystallized region (Montgomery & Winget 1999).

BPM37093 (Kanaan et al. 2005, e.g.) is the only previously known high mass pulsator, with spectroscopic and seismological values around $1\,M_\odot$. At the temperatures around the ZZ Ceti instability strip, stars with such mass are expected to be substantially crystallized. For BPM37093, the best models predict that 66–92% of the star is crystallized (Kanaan et al. 2005). This star has been observed for many seasons, showing strong amplitude variation. McGraw (1976), for example, reported it to be non-variable, while Kanaan et al. (1998) reported that on one occasion, all the previously observed pulsation modes vanished below their detection limit of 1 mma. Because we do not know the timescales for the phase when the amplitudes are below this limit, even a high mass NOV1 can be a pulsator. Besides, these stars are expected, in general, to have smaller pulsation amplitudes than more average mass white dwarfs due to the size of the resonant cavity being reduced by higher mass (i.e., smaller stellar radius) and/or by the crystallized mass of the star.

In the case of SDSS J005047.60-002316.9, the spectroscopic mass is also above $1\,M_\odot$, so at the observed temperature, its core should also be crystallized. It was classified as NOV6 by Mukadam et al. (2004a). We have observed this star for many seasons, since 2005, with a 1/1000 false alarm probability of 3.73 mma. Combining the two runs in 2007 (see Table 1), the false alarm probability for the 3.5 mma at 584 s (Scargle 1982) is 1/250. Pulsations, if present, might be damped by the proposed crystallized core. Because pulsations can be used to determine the crystallized fraction, which in turn can help determine the C/O fraction in the core, uncertain theoretically because of the large uncertainty on the $C(\alpha,\gamma)O$ cross section, we encourage further observations of SDSS J005047.60-002316.9 to determine if the star is really a variable.

## 7 CONCLUSIONS AND FINAL REMARKS

In this paper, we have presented a report on our latest searches for ZZ Ceti stars. We have discovered ten new pulsators and lowered the detection limit for ten stars with spectroscopic temperatures close or within the boundaries of the instability strip.

For the new ZZ Ceti stars, we have done a first seismological study, even though there are not many data available other than the discovery and confirmation runs. In the future, with the monitoring of these stars, it is likely that more modes will be revealed, allowing a detailed seismological study. The masses from seismology are consistent with those determined from the SDSS spectra, but the temperatures from seismology are on average cooler than from the spectra, specially for the cooler stars.

There are still NOVs within the ZZ Ceti instability strip, but until their variability limits can be reduced to of order 1 mma (the lowest amplitude known pulsator), we cannot truly declare them pulsators or non-pulsators. Lowering the detection limit to much below 1 mma becomes problematic as we try to decide what level of variability constitutes unstable pulsators and what level is simply normal stellar variability. We should also re-observe stars outside the instability strip, which are reported as NOVs, because their detection limits need to be lowered to 1 mma. Our results imply that we cannot claim we know the edges of the ZZ Ceti instability strip, until we lower the detection limit for all NOVs in the nearby temperature range.

**ACKNOWLEDGMENTS**

We acknowledge support from the CNPq-Brazil.

The data was acquired at SOAR on proposals: SO05A, SO05B, SO06A, SO06B, SO07A-021, SO07B-018, TR07B-003, SO08A-015, SO08B-012, SO09A-012, and TR09B-015.

Funding for the Sloan Digital Sky Survey (SDSS) and SDSS-II has been provided by the Alfred P. Sloan Foundation, the Participating Institutions, the National Science Foundation, the U.S. Department of Energy, the National Aeronautics and Space Administration, the Japanese Monbukagakusho, and the Max Planck Society, and the Higher Education Funding Council for England. The SDSS Web site is http://www.sdss.org/.

The SDSS is managed by the Astrophysical Research Consortium (ARC) for the Participating Institutions. The Participating Institutions are the American Museum of Natural History, Astrophysical Institute Potsdam, University of Basel, University of Cambridge, Case Western Reserve University, The University of Chicago, Drexel University, Fermilab, the Institute for Advanced Study, the Japan Participation Group, The Johns Hopkins University, the Joint Institute for Nuclear Astrophysics, the Kavli Institute for Particle Astrophysics and Cosmology, the Korean Scientist Group, the Chinese Academy of Sciences (LAMOST), Los Alamos National Laboratory, the Max-Planck-Institute for Astronomy (MPIA), the Max-Planck-Institute for Astrophysics (MPA), New Mexico State University, Ohio State University, University of Pittsburgh, University of Portsmouth, Princeton University, the United States Naval Observatory, and the University of Washington.